\documentclass[prl,aps,twocolumn,amssymb,amsmath]{revtex4-2}

\usepackage{amsmath,amssymb,mathtools,bm}
\usepackage[colorlinks=true, pdfstartview=FitV, linkcolor=red, citecolor=blue, urlcolor=blue]{hyperref}
\usepackage{hyperref}
\usepackage{microtype}

\newcommand{\tr}{\operatorname{tr}}
\newcommand{\dd}{\mathrm d}
\newcommand{\ii}{\mathrm i}
\newcommand{\ee}{\mathrm e}
\newcommand{\U}{\mathrm{U}}
\newcommand{\SU}{\mathrm{SU}}
\newcommand{\SO}{\mathrm{SO}}
\newcommand\sect[1]{{\it #1}---}

\begin{document}
\title{Spin-One Secondary Pairing in Two-Flavor Color Superconductivity:\\
Spinful Relativistic Superfluidity and Anomaly Matching}
\author{Naoki Yamamoto}
\affiliation{Department of Physics, Keio University, Yokohama 223-8522, Japan}

\begin{abstract}
We study secondary pairing in the two-flavor color-superconducting (2SC) phase, where the residual ungapped quarks form a same-chirality $J^P=1^+$ condensate driven by an attractive instanton-induced interaction. We determine its symmetry realization, quasiparticle structure, anomaly matching, and low-energy effective theory. The pairing gap is necessarily nodal; in particular, the complex axial state has two point nodes and realizes a spinful relativistic superfluid. This state preserves the full chiral symmetry $\SU(2)_{\rm L}\times\SU(2)_{\rm R}$ while breaking the modified baryon-number symmetry $\U(1)_{\widetilde{\rm B}}$ and spatial rotations, with rotations about the nodal axis locked to the condensate phase. This locking gives rise to a Berry term, the Mermin--Ho relation, and a type-B orientational Nambu--Goldstone mode in addition to the superfluid phonon. We also show how anomaly matching is reorganized by secondary pairing: the perturbative mixed $\SU(2)_{\rm L,R}^2\U(1)_{\widetilde{\rm B}}$ anomaly is realized by a Wess--Zumino coupling of the superfluid phonon, whereas the $\SU(2)_{\rm L}$ and $\SU(2)_{\rm R}$ Witten anomalies are carried by the point-node Bogoliubov--de Gennes flavor doublets. The resulting theory provides a concrete dense-QCD realization of a spinful relativistic superfluid with nodal fermions required by anomaly matching.
\end{abstract}

\maketitle

\sect{Introduction}%
Quantum chromodynamics (QCD) at high density exhibits color superconducting phases \cite{Alford:2007xm}. The two-flavor color-superconducting (2SC) phase arises from the leading BCS instability of massless $N_{\rm f}=2$ QCD at large quark chemical potential \cite{Bailin:1983bm,Alford:1997zt,Rapp:1997zu}. In a convenient gauge, the red and green quarks form the color-antitriplet, flavor-singlet, spin-zero condensate, whereas the blue up and down quarks remain ungapped. 
Secondary pairing of these residual quarks has also been considered since the early literature on color superconductivity \cite{Alford:1997zt,Buballa:2002wy,Alford:2002rz}. Since a pair of two blue quarks is necessarily in the symmetric color-sextet representation, the perturbative one-gluon-exchange interaction does not provide an attractive channel. 
On the other hand, the instanton-induced interaction acts as an attraction in the flavor-singlet $J^P=1^+$ spin-one same-chirality $C\sigma_{0i}$ channel~\cite{Alford:2002rz}. 
A key unresolved conceptual question is how anomaly matching is maintained once the residual quarks in conventional 2SC themselves undergo secondary pairing \cite{Alford:2001dt}. 

In this paper, we connect the physics of secondary pairing in 2SC with anomaly matching and spinful relativistic superfluidity. We determine the symmetry realization, quasiparticle structure, anomaly matching, and low-energy effective field theory (EFT) of the same-chirality $J^P=1^+$ secondary-paired phase. We first show that the quasiparticle spectrum is necessarily nodal in the massless limit. As a controlled weak-coupling reference, the quartic Ginzburg--Landau (GL) term favors a complex axial state with two point nodes. This axial state preserves the full chiral symmetry while spontaneously breaking the modified baryon-number symmetry and spatial rotations, with rotations about the nodal axis locked to the condensate phase. We further show that this phase provides a concrete dense-QCD realization of a spinful relativistic superfluid of Ref.~\cite{Son:2026tcm}, characterized by a phase-orientation topological interaction, the Mermin--Ho relation, and a type-B orientational Nambu--Goldstone (NG) mode. At the same time, this EFT satisfies the anomaly matching constraint: the resulting superfluid phonon realizes the perturbative mixed anomaly locally, whereas the nodal Bogoliubov--de Gennes (BdG) flavor doublets provide the infrared (IR) carriers of the Witten $\mathbb Z_2$ anomalies. 
We thereby resolve how the anomalies carried by the residual quarks in ordinary 2SC survive after those very quasiparticles undergo secondary pairing. 
Throughout this paper, $i,j,k=1,2,3$ denote spatial indices.

\sect{2SC phase}%
We consider massless two-flavor QCD with equal up- and down-quark chemical potentials, $\mu_{\rm u}=\mu_{\rm d}\eqqcolon\mu$, at zero temperature.
We turn off electromagnetic and weak interactions. 
The continuous global symmetry of the system is $G=\SU(2)_{\rm L}\times \SU(2)_{\rm R}\times \U(1)_{\rm B}$.
The anomalous $\U(1)_{\rm A}$ symmetry is irrelevant to the following discussion; for the associated $\eta$ sector, see Refs.~\cite{Son:2000fh,Nishimura:2024odm}.

At high density, the conventional 2SC condensate due to the primary pairing instability takes the form \cite{Bailin:1983bm,Alford:1997zt,Rapp:1997zu} 
\begin{equation}
  \Phi^\alpha_{\rm 2SC} \coloneqq \epsilon^{\alpha\beta\gamma}\epsilon^{AB}
  \left\langle q_{\beta A}^{\rm T}C\gamma_5q_{\gamma B} \right\rangle,
  \label{eq:2SC-order-parameter}
\end{equation}
where $\alpha,\beta,\gamma=1,2,3$ are color indices and $A,B=1,2$ are flavor indices. The condensate transforms as a positive-parity, spin-zero, flavor-singlet color-antitriplet pair.
Using a color gauge transformation, the condensate can be oriented along the third color direction, 
$\Phi^\alpha_{\rm 2SC} \propto \Delta_{\rm 2SC}\delta^{\alpha3}$.
In this gauge, quarks of colors $1$ and $2$ (red and green) are paired, while quarks of color $3$ (blue) remain ungapped at the primary 2SC level. 

The 2SC condensate is a singlet under both $\SU(2)_{\rm L}$ and $\SU(2)_{\rm R}$, and does not break the chiral symmetry. 
The color gauge group is Higgsed from $\SU(3)_{\rm c}$ to $\SU(2)_{\rm c}$, while a combination of the ordinary baryon-number generator and the broken color generator remains as an unbroken global symmetry.
With $B=\frac{1}{3}\operatorname{diag}(1,1,1)$ and $T_8=\frac{1}{2\sqrt3}\operatorname{diag}(1,1,-2)$, its generator may be chosen as \cite{Schafer:1999pb}
\begin{equation}
  \widetilde B = B-\frac{2}{\sqrt3}T_8 = \operatorname{diag}(0,0,1).
\end{equation}
The unbroken continuous global symmetry of the conventional 2SC phase is thus $\SU(2)_{\rm L}\times \SU(2)_{\rm R}\times \U(1)_{\widetilde {\rm B}}$.
The residual blue up and down quarks form one left-handed and one right-handed flavor doublet, each carrying $\widetilde B=1$, and constitute the gapless fermionic degrees of freedom.

\sect{Anomalies carried by the residual quarks}%
For perturbative anomalies in four spacetime dimensions (4D), the anomalous variation of the generating functional can be encoded compactly in a gauge-invariant closed six-form anomaly polynomial $I_6$. The relation to the 4D anomaly is given by the descent construction \cite{Zumino:1983rz}, $I_6=\dd I_5^{(0)}$ and $\delta I_5^{(0)}=\dd I_4^{(1)}$, so that the anomalous variation of the 4D generating functional on a closed four-manifold $M_4$ is
\begin{equation}
  \delta W = -2\pi\int_{M_4} I_4^{(1)}.
\end{equation}
Here, the subscript denotes the differential-form degree, while the superscript indicates the order in the gauge-transformation parameter. 

We now apply this framework to the global symmetry of the ordinary 2SC phase. We introduce background gauge fields $A_{\rm L}$ and $A_{\rm R}$ for $\SU(2)_{\rm L}$ and $\SU(2)_{\rm R}$, respectively, and $b$ for $\U(1)_{\widetilde{\rm B}}$. Their field strengths are $F_\chi=\dd A_\chi-\ii A_\chi\wedge A_\chi$ ($\chi={\rm L,R}$) and $F_{\tilde{\rm B}}=\dd b$. We use the integral characteristic classes
\begin{equation}
  c_1(b) = \frac{F_{\tilde {\rm B}}}{2\pi}, 
  \qquad
  c_2(F_\chi) = \frac{1}{8\pi^2}\tr\left(F_\chi\wedge F_\chi\right),
\end{equation}
where, in the fundamental representation, the $\SU(2)$ generators $T^a$ ($a=1,2,3$) are normalized as $\tr(T^aT^b)=\frac12\delta^{ab}$, so that $\int_{M_4} c_2(F_\chi)\in\mathbb Z$.

With these conventions, the perturbative mixed anomaly of the ordinary 2SC phase is
\begin{equation}
  I_6^{\rm pert} = c_1(b)\left[c_2(F_{\rm L})-c_2(F_{\rm R})\right].
  \label{eq:I6}
\end{equation}
This encodes the $\SU(2)_{\rm L}^2\U(1)_{\widetilde {\rm B}}$ and $\SU(2)_{\rm R}^2\U(1)_{\widetilde {\rm B}}$ anomalies, with opposite signs for the two chiralities. 
For example, the coefficient of the $\SU(2)_{\rm L}^2\U(1)_{\rm B}$ anomaly in the UV is 
$3\times\frac13\times\frac12 = \frac12$, 
where the factors arise from the number of colors, the quark baryon charge, and the $\SU(2)$ group trace, respectively. 
In the ordinary 2SC phase, the single residual flavor doublet in the IR has $\widetilde B=1$ and gives 
$1\times1\times\frac12=\frac12$.
Therefore, the mixed anomaly is exactly matched between UV QCD and the IR theory of the 2SC phase \cite{Sannino:2000kg}. 
Note that, for $N_{\rm f}=2$, the perturbative $\SU(2)^3$ anomaly vanishes because $d^{abc}\coloneqq2\tr[T^a\{T^b,T^c\}]=0$. 

In addition, two-flavor QCD has a genuinely global anomaly that has no perturbative anomaly-polynomial representation: the Witten $\mathbb Z_2$ anomaly of each chiral $\SU(2)$ factor \cite{Witten:1982fp}. Indeed, the UV theory (two-flavor and three-color QCD) contains three $\SU(2)_{\rm L}$ doublets and three $\SU(2)_{\rm R}$ doublets. Since $3=1\pmod 2$, each chiral $\SU(2)$ has a nontrivial Witten anomaly. In the ordinary 2SC phase, the single gapless blue quark doublet in each chiral sector reproduces the corresponding global anomaly.

\sect{Pairing of the residual quarks}%
The gapless residual blue quarks of ordinary 2SC may in principle undergo secondary pairing \cite{Alford:1997zt,Buballa:2002wy,Alford:2002rz}. Because both quarks have the same fixed color, their pair belongs to the symmetric color-sextet representation. 
For the one-color, flavor-singlet $J^P=1^+$ condensate $\left\langle q^{\rm T}\ii\tau_2 C\sigma_{03}q \right\rangle \neq0$, the instanton-induced interaction is attractive \cite{Buballa:2002wy,Alford:2002rz}.
This is the relativistic $J^P=1^+$ operator, which takes a simple form after projection onto positive-energy massless quarks. Suppressing the fixed blue color index, the same-chirality two-component order parameters may be written as
\begin{equation}
  \Phi_\chi^i = \epsilon^{AB} \left\langle q_{{\chi},A}^{\rm T} \ii\sigma^2\sigma^i q_{{\chi},B} \right\rangle,
  \qquad
  \chi={\rm L,R}.
  \label{eq:PhiLR}
\end{equation}
We parametrize the condensates as $\Phi_\chi^i=\Phi_\chi d_\chi^i$, with $\bm d_\chi^\dagger\bm d_\chi=1$.
For a parity-invariant state, the left- and right-handed condensates can be chosen with the same amplitude and spatial orientation. 
Writing the corresponding gap amplitudes as $\Delta_\chi$, below we set $\Delta_{\rm L}=\Delta_{\rm R}\eqqcolon\Delta\geq0$ and ${\bm d}_{\rm L}={\bm d}_{\rm R}\eqqcolon{\bm d}$.

The helicity projection implies that the magnitude of the quasiparticle gap has an angular dependence proportional to $|\bm d\cdot\hat{\bm p}|$.
We define the normalized angular form factor
\begin{equation}
  f(\hat{\bm p}) = \sqrt{3}\,\bm d\cdot\hat{\bm p},
  \qquad
  \left\langle |f(\hat{\bm p})|^2\right\rangle=1, 
  \label{eq:vector-formfactor}
\end{equation}
where $\langle\cdots\rangle$ denotes the average over the Fermi surface. 
The momentum-dependent gap function is $\Delta(\hat{\bm p})=\Delta f(\hat{\bm p})$, with magnitude $|\Delta(\hat{\bm p})|=\Delta|f(\hat{\bm p})|$.
The corresponding quasiparticle spectrum is
\begin{equation}
  E_\chi(\bm p) = \sqrt{\xi_{\bm p}^2 +\Delta^2 |f(\hat{\bm p})|^2},
  \qquad
  \xi_{\bm p}=|{\bm p}|-\mu.
  \label{eq:spectrum}
\end{equation}
For the polar representative $C\sigma_{03}$, corresponding to $\bm d=\hat{\bm z}$, one has $|f(\hat{\bm p})|=\sqrt3|\cos\theta|$, yielding a line node at the equator in the massless limit, in agreement with the relativistic dispersion found in Ref.~\cite{Alford:2002rz}.

Generally, the spectrum is necessarily nodal for any complex vector $\bm d$. Writing $\bm d=\bm a+\ii\bm b$ with real vectors $\bm a$ and $\bm b$, a node requires $\bm a\cdot\hat{\bm p}=\bm b\cdot\hat{\bm p}=0$. If $\bm a$ and $\bm b$ are linearly independent, the two solutions are $\hat{\bm p}=\pm(\bm a\times\bm b)/|\bm a\times\bm b|$, corresponding to a pair of antipodal point nodes; if they are parallel, the nodes form a line on the Fermi surface. Hence, a flavor-singlet $J^P=1^+$ vector condensate cannot fully gap the massless residual quarks.

\sect{Ginzburg--Landau analysis}%
The orientation of the complex vector $\bm d$ is degenerate at quadratic order by rotational symmetry, but the quartic GL term distinguishes different nodal structures. For a generally anisotropic gap $\Delta f(\hat{\bm p})$, weak-coupling BCS theory gives quadratic and quartic GL coefficients proportional to $\langle |f(\hat{\bm p})|^2\rangle$ and $\langle |f(\hat{\bm p})|^4\rangle$, respectively \cite{GorkovMelikBarkhudarov1964,Leggett1975}. 
The GL free energy can then be written as \footnote{Although we otherwise work at zero temperature, the GL analysis is understood near the critical temperature, where the expansion is controlled; the subsequent low-energy analysis is performed at zero temperature.} 
\begin{equation}
  \Omega = \alpha\Delta^2 + \frac{\beta_0}{2} K[\bm d]\Delta^4 +\cdots,
  \qquad
  \beta_0>0,
  \label{eq:GL}
\end{equation}
where $K[\bm d] \coloneqq \left\langle |f(\hat{\bm p})|^4\right\rangle$.
Using Eq.~\eqref{eq:vector-formfactor} and
\begin{equation}
  \left\langle \hat p_i\hat p_j\hat p_k\hat p_l \right\rangle
  =\frac{1}{15} \left(\delta_{ij}\delta_{kl}+\delta_{ik}\delta_{jl}+\delta_{il}\delta_{jk}\right),
\end{equation}
one obtains
\begin{equation}
  K[\bm d]=\frac35\left(2+|\bm d\cdot\bm d|^2\right).
  \label{eq:K}
\end{equation}
The minimum is $K_{\rm axial}=6/5$ for $\bm d\cdot\bm d=0$. Up to a spatial rotation and an overall phase, a representative is
\begin{equation}
  \bm d=\frac{1}{\sqrt2}\left(\hat{\bm e}_1+\ii\hat{\bm e}_2 \right), 
  \qquad
  f_{\rm axial}(\hat{\bm p})=\sqrt{\frac32}\sin\theta\,\ee^{\ii\phi},
  \label{eq:axial-formfactor}
\end{equation}
where we choose $\hat{\bm e}_{1}=\hat{\bm x}$ and $\hat{\bm e}_{2}=\hat{\bm y}$.
This complex axial state has two point nodes at the north and south poles 
\footnote{Here, ``polar'' and ``axial'' correspond to $m_J=0$ and $m_J=\pm1$, respectively, where $m_J$ denotes the projection of the $J=1$ Cooper-pair angular momentum onto the $z$ axis. This terminology should not be confused with the axial symmetry of QCD.}. 
By comparison, the real polar state $\bm d=\hat{\bm z}$ has $K_{\rm polar}=9/5$ and a line node. For $\alpha<0$, minimizing Eq.~\eqref{eq:GL} gives $\Omega_{\rm min}=-\alpha^2/(2\beta_0K)$, so the axial state gains more condensation energy than the polar state within this weak-coupling GL analysis. In the following, we focus on the axial state. However, the internal-symmetry realization and anomaly matching are independent of this choice of spatial pairing structure.

\sect{Symmetry realization and perturbative mixed anomaly}%
We now study the symmetry realization and anomaly matching of the secondary-paired 2SC phase.
Because the order parameter in Eq.~\eqref{eq:PhiLR} is a flavor singlet for each chirality, the full chiral symmetry $\SU(2)_{\rm L}\times\SU(2)_{\rm R}$ remains unbroken. Each residual blue quark has $\widetilde B=1$, so the blue-blue condensate carries $\widetilde B=2$ and breaks $\U(1)_{\widetilde{\rm B}}$ down to $\mathbb Z_2$. 
The spontaneous breaking of $\U(1)_{\widetilde{\rm B}}$ gives rise to an NG mode, the superfluid phonon $\varphi$, which transforms as $\varphi\rightarrow\varphi+2\alpha_{\widetilde{\rm B}}$ under $\U(1)_{\widetilde{\rm B}}$.
The vector order parameter also breaks spatial rotations. For the axial state in Eq.~\eqref{eq:axial-formfactor}, rotations about the nodal axis act as a phase rotation of $\bm d$ and are locked to the condensate phase. 
The associated orientational NG modes are spectators for the internal 't Hooft anomalies considered below.

The perturbative mixed anomaly in Eq.~\eqref{eq:I6} can be represented locally by the superfluid phonon. Since $\delta\varphi=2\alpha_{\widetilde{\rm B}}$ under a $\U(1)_{\widetilde{\rm B}}$ transformation, the Wess--Zumino (WZ) coupling
\begin{equation}
  S_{\rm WZ}
  =-\frac12\int_{M_4} \varphi \left[c_2(F_{\rm L})-c_2(F_{\rm R})\right]
  \label{eq:phononWZ}
\end{equation}
reproduces the perturbative $\SU(2)_{\rm L,R}^2\U(1)_{\widetilde{\rm B}}$ anomaly.  
Because $\varphi\sim\varphi+2\pi$, a $2\pi$ shift changes the exponentiated action by
\begin{align}
  \ee^{\ii S_{\rm WZ}}
  \rightarrow
  \ee^{\ii S_{\rm WZ}}(-1)^{k_{\rm L}-k_{\rm R}}, 
  \qquad
  k_\chi=\int_{M_4}c_2(F_\chi).
\end{align}
The factor $1/2$, required locally by the charge-2 transformation of $\varphi$, implies that a $2\pi$ shift of $\varphi$ changes the exponentiated action by a sign whenever $k_{\rm L}-k_{\rm R}$ is odd. Hence, the local phonon WZ term by itself is not globally well defined; the required mod-2 completion is naturally expected to be supplied by the quasiparticle Witten anomalies discussed below.

\sect{Nodal BdG quasiparticles and Witten anomalies}%
Since the $J^P=1^+$ condensate is a flavor singlet and leaves the full chiral symmetry unbroken, there is no chiral NG sector that could realize the $\SU(2)_{\rm L}$ and $\SU(2)_{\rm R}$ Witten anomalies. An IR phase that is simultaneously fully gapped, chirally symmetric, and topologically trivial would be inconsistent with the UV Witten anomalies. However, the $J^P=1^+$ state avoids this obstruction because its quasiparticle spectrum is necessarily nodal.

The left-handed blue quark flavor doublet is denoted by $(q_{\rm L})_A$, which transforms as $(q_{\rm L})_A \rightarrow (U_{\rm L})_A{}^B(q_{\rm L})_B$ under $U_{\rm L}\in\SU(2)_{\rm L}$. 
Introducing $\epsilon_{AB}=(\ii\tau_2)_{AB}$ and the conjugate doublet $(\widetilde q_{\rm L})_A\coloneqq\epsilon_{AB}(q_{\rm L}^*)_B$, pseudoreality of the $\SU(2)$ fundamental representation implies
$(\widetilde q_{\rm L})_A\rightarrow(U_{\rm L})_A{}^B(\widetilde q_{\rm L})_B$.
Hence, both particle and hole components transform as fundamentals of the same $\SU(2)_{\rm L}$.
For the axial state, the two point nodes lie at $\bm p_\star=\mu\hat{\bm n}_\star$ and $-\bm p_\star$. The slowly varying left-handed patch fields are denoted by $q_{{\rm L},+}$ and $q_{{\rm L},-}$ and form the Nambu spinor $(\Psi_{\rm L})_A=\left((q_{{\rm L},+})_A,\,\epsilon_{AB}(q_{{\rm L},-}^*)_B \right)^{\rm T}$.

We write $\bm p=\bm p_\star+\bm k$ and choose an orthonormal local basis $\{\hat{\bm n}_\star,\hat{\bm e}_1,\hat{\bm e}_2\}$. The residual momentum can be decomposed as $\bm k=k_\parallel\hat{\bm n}_\star+k_1\hat{\bm e}_1+k_2\hat{\bm e}_2$.
For a background $\SU(2)_{\rm L}$ gauge field $A_{{\rm L}\mu}$, we define ${\cal D}_\mu=\partial_\mu-\ii A_{{\rm L}\mu}$, ${\cal D}_\parallel\coloneqq\hat n_\star^i{\cal D}_i$, and ${\cal D}_a\coloneqq\hat e_a^i{\cal D}_i$ ($a=1,2$).
Expanding Eq.~\eqref{eq:axial-formfactor} near a point node gives
\begin{equation}
  \Delta f_{\rm axial}(\hat{\bm p})\simeq v_\Delta(k_1+\ii k_2),
  \qquad
  v_\Delta=\sqrt{\frac32}\frac{\Delta}{\mu}.
\end{equation}
At the same time, $\xi_{\bm p}\simeq k_\parallel$. 
At vanishing phase and orientational gradients, the low-energy BdG Lagrangian is then
\begin{align}
  {\cal L}^{(0)}_{\rm node,L}
  &=\Psi_{\rm L}^\dagger\big[\ii {\cal D}_0 -\tau_3^{\rm N}(-\ii {\cal D}_\parallel)
    \nonumber\\
  &\qquad
    -v_\Delta\left\{\tau_1^{\rm N}(-\ii {\cal D}_1)+\tau_2^{\rm N}(-\ii {\cal D}_2)\right\}\big]\Psi_{\rm L}+\cdots
  \label{eq:nodal-BdG}
\end{align}
and describes an emergent anisotropic Weyl-like quasiparticle. 
Here, $\tau_i^{\rm N}$ denotes the Pauli matrices acting in Nambu space to distinguish them from flavor Pauli matrices.
The right-handed sector is obtained similarly, with $\SU(2)_{\rm L}$ replaced by $\SU(2)_{\rm R}$.

Equation~\eqref{eq:nodal-BdG} shows that the nodal BdG quasiparticles transform as an $\SU(2)_{\rm L}$ flavor doublet, just as the gapless blue quarks in ordinary 2SC. For $v_\Delta \neq 0$, the theory of anisotropic Weyl-like fermions can be continuously deformed into that of isotropic Weyl fermions without changing the anomaly. Hence, it carries the same anomaly as a single $\SU(2)_{\rm L}$ Weyl doublet. Although the gap has two antipodal nodes, they are related by the BdG particle-hole redundancy and do not represent two independent Weyl doublets. Together with the analogous right-handed sector, the nodal quasiparticles realize the $\SU(2)_{\rm L}$ and $\SU(2)_{\rm R}$ Witten anomalies.

\sect{Phase-orientation coupling and low-energy EFT}%
The spontaneous breaking of spatial rotations also leads to the orientational NG modes. 
In the axial state, however, the superfluid and orientational sectors are not independent. 
We promote the fixed reference vectors $\hat{\bm e}_{1,2}$ to a local orthonormal frame $\bm e_{1,2}(x)$ describing the orientational fluctuations, and define 
$\bm e_+ \coloneqq \frac{1}{\sqrt2} \left(\bm e_1+\ii\bm e_2 \right)$, $\bm\ell \coloneqq \bm e_1\times\bm e_2=-\ii\bm e_+^*\times\bm e_+$, 
so that $\{\bm e_1,\bm e_2,\bm\ell\}$ forms a local orthonormal triad. 
Since the helicity-projected gap is proportional to $|\bm d\cdot\hat{\bm p}|$, it vanishes when $\hat{\bm p}$ is orthogonal to both $\bm e_1$ and $\bm e_2$, namely, at $\hat{\bm p}=\pm\bm\ell$.
Hence, $\bm\ell$ defines the nodal axis connecting the two point nodes. Writing the order parameter as $\Phi^i\propto \ee^{\ii\varphi}e_+^i$, this parametrization has the local redundancy
\begin{equation}
  \bm e_+(x)\rightarrow\ee^{\ii\beta(x)}\bm e_+(x),
  \qquad
  \varphi(x)\rightarrow\varphi(x)-\beta(x),
  \label{eq:axial-frame-redundancy}
\end{equation}
which reflects the locking between rotations about $\bm\ell$ and $\varphi$. Correspondingly, the continuous symmetry breaking pattern in the $\U(1)_{\widetilde{\rm B}}$ and rotational sectors is 
$\U(1)_{\widetilde{\rm B}}\times\SO(3)_J\rightarrow\U(1)_{\rm locked}$. 

Because the choice of the transverse basis $(\bm e_1,\bm e_2)$ for a given $\bm\ell$ is defined only up to a local rotation about $\bm\ell$, it has a local $\SO(2)\simeq\U(1)$ frame redundancy. The associated composite connection is $a_\mu \coloneqq -\ii\bm e_+^*\cdot\partial_\mu\bm e_+$, which transforms as 
$a_\mu\rightarrow a_\mu+\partial_\mu\beta$
under Eq.~\eqref{eq:axial-frame-redundancy}. Since $D_\mu\varphi=\partial_\mu\varphi-2b_\mu$, the combination 
${\cal V}_\mu \coloneqq \frac12\left( D_\mu\varphi+a_\mu\right)$
is invariant under both the background $\U(1)_{\widetilde{\rm B}}$ transformation and the local frame redundancy. 
The factor $1/2$ reflects $\widetilde B=2$ of the Cooper pair.

The curvature of the orientational connection is determined by the texture of the nodal axis,
$\partial_\mu a_\nu-\partial_\nu a_\mu=\bm\ell\cdot\left(\partial_\mu\bm\ell\times \partial_\nu\bm\ell\right)$,
and therefore
\begin{equation}
  \partial_\mu{\cal V}_\nu - \partial_\nu{\cal V}_\mu=-F^{(\widetilde{\rm B})}_{\mu\nu}+\frac12\bm\ell\cdot\left(\partial_\mu\bm\ell \times \partial_\nu\bm\ell\right),
  \label{eq:Mermin-Ho}
\end{equation}
where $F^{(\widetilde{\rm B})}_{\mu\nu}
  = \partial_\mu b_\nu-\partial_\nu b_\mu$.
Without $b_{\mu}$, Eq.~\eqref{eq:Mermin-Ho} gives a 4D extension of the Mermin--Ho relation: a smooth orientational texture can carry superfluid vorticity \cite{MerminHo1976,Son:2026tcm}.

A local derivative expansion of the bosonic EFT can be organized in terms of ${\cal V}_\mu$ and $\bm\ell$. A minimal set of leading terms relevant for the present discussion is
\begin{align}
  {\cal L}_{\rm boson}
  ={}&n_{\widetilde{\rm B}}{\cal V}_0+\frac{f_{\varphi}^2}{2} \! \left[{\cal V}_0^2-v_{\parallel}^2(\ell^i{\cal V}_i)^2-v_{\perp}^2(\delta^{ij}-\ell^i\ell^j){\cal V}_i{\cal V}_j \right]
  \nonumber\\
  &-\frac{\kappa_\parallel}{2}\left(\ell^i\partial_i\bm\ell\right)^2-\frac{\kappa_\perp}{2}\left(\delta^{ij}-\ell^i\ell^j\right)\partial_i\bm\ell\cdot\partial_j\bm\ell+\cdots.
  \label{eq:bosonic-EFT}
\end{align}
Here, $n_{\widetilde{\rm B}}$ is the equilibrium modified baryon density (equal to the ordinary baryon density $n_{\rm B}$ in color-neutral 2SC matter), while $f_\varphi$, $v_{\parallel,\perp}$, and $\kappa_{\parallel,\perp}$ are low-energy constants.
The coefficient of the term linear in ${\cal V}_\mu$ is fixed by the conserved current: at leading order, one has $j_{\widetilde{\rm B}}^\mu{\cal V}_\mu$, which reduces to $n_{\widetilde{\rm B}}{\cal V}_0$ in the local rest frame. Using ${\cal V}_0=\frac12(D_0\varphi+a_0)$, the $\partial_0\varphi$ contribution is a total derivative, whereas the $a_0$ contribution gives the Berry term for the orientational mode.

To see its geometric meaning explicitly, we write
$\bm\ell=(\sin\theta\cos\phi,\sin\theta\sin\phi,\cos\theta)$.
A local orthonormal frame $(\bm e_1,\bm e_2)$ transverse to $\bm\ell$, chosen to be regular near the north pole, gives
$a_0=(1-\cos\theta)\dot\phi$. 
For a closed trajectory $C$ traced by $\bm\ell$ on $S^2$,
\begin{equation}
  \oint_C a
  = \oint_C (1-\cos\theta)\,\dd\phi
  = \int_{\Sigma_C} \sin\theta\,\dd\theta\,\dd\phi
  = \Omega[C],
  \label{eq:Berry-solid-angle}
\end{equation}
where $\Sigma_C\subset S^2$ is an oriented surface with $\partial\Sigma_C=C$, and $\Omega[C]$ is its oriented solid angle.
Since the $a_0$ part of the first term in Eq.~\eqref{eq:bosonic-EFT} is $\frac12n_{\widetilde{\rm B}}a_0$, the Berry contribution to the action for a spatially uniform configuration in a volume $V$ is
\begin{equation}
  \frac{S_{\rm B}}{V}={\cal J}\,\Omega[C],
  \qquad
  {\cal J} = \frac{n_{\widetilde{\rm B}}}{2}.
  \label{eq:J-Berry}
\end{equation}
This is the Berry phase associated with the orientational degrees of freedom. The coefficient ${\cal J}$ is thus identified as the macroscopic angular-momentum density along $\bm\ell$. The same result follows directly from the microscopic quantum numbers: each axial Cooper pair carries angular momentum $J_{\bm\ell}=1$ along $\bm\ell$ and $\widetilde B=2$, so that the angular momentum per unit modified baryon charge is $s=1/2$, and ${\cal J}=s n_{\widetilde{\rm B}}$ at zero temperature. This agrees with the general spinful-superfluid relation of Ref.~\cite{Son:2026tcm}. See End Matter for a topological formulation of the coupling between the superfluid and orientational sectors. 

The resulting NG structure can be found by expanding Eq.~\eqref{eq:bosonic-EFT} around $\bm\ell=\hat{\bm z}$ as
$\bm\ell=\left(\pi_1,\,\pi_2,\,\sqrt{1-\pi_1^2-\pi_2^2}\right)$.
Using the corresponding smooth transverse frame obtained by the minimal rotation that maps $\hat{\bm z}$ to $\bm\ell$, the composite connection takes the form
\begin{equation}
  a_\mu=\frac{\pi_1\partial_\mu\pi_2-\pi_2\partial_\mu\pi_1}{1+\sqrt{1-\pi_1^2-\pi_2^2}}.
\end{equation}
Expanding around $\bm\pi=0$, the quadratic Lagrangian for the orientational sector is
\begin{equation}
  {\cal L}_{\rm rot}^{(2)}=\frac{{\cal J}}{2}\left(\pi_1\dot\pi_2-\pi_2\dot\pi_1 \right)-\frac{\kappa_\parallel}{2}(\partial_\parallel\bm\pi)^2- \frac{\kappa_\perp}{2}(\bm\nabla_\perp\bm\pi)^2+\cdots.
  \label{eq:rotational-NG-EFT}
\end{equation}
The two broken transverse rotations form a canonically conjugate pair and give a single type-B NG mode \cite{Watanabe:2012hr, Hidaka:2012ym} with the dispersion relation (see also Ref.~\cite{Pang:2010wk})
\begin{equation}
  \omega=\frac{\kappa_\parallel k_\parallel^2+\kappa_\perp k_\perp^2}{{\cal J}}+\cdots.
  \label{eq:rotational-NG-dispersion}
\end{equation}
Together with the superfluid phonon, the bosonic sector thus contains one type-A and one type-B NG mode.

The orientational modes also couple directly to the nodal quasiparticles. For a slowly varying texture, the fixed local basis used above is promoted according to
$\hat{\bm n}_\star \rightarrow \bm\ell(x)$ and $\hat{\bm e}_a \rightarrow \bm e_a(x)$, so that ${\cal D}_\parallel \rightarrow \ell^i{\cal D}_i$ and ${\cal D}_a \rightarrow e_a^i{\cal D}_i$ $(a=1,2)$. Consequently, the texture shifts the nodal positions, $\bm p_\star(x)=\pm\mu\bm\ell(x)$, and rotates the local anisotropic Weyl cone. 

Under a $\U(1)_{\widetilde{\rm B}}$ transformation, $\Psi_{\rm L}\rightarrow \ee^{\ii\alpha_{\widetilde{\rm B}}\tau_3^{\rm N}}\Psi_{\rm L}$ and $\varphi\rightarrow\varphi+2\alpha_{\widetilde{\rm B}}$, so that $\ee^{-\ii\varphi\tau_3^{\rm N}/2}\Psi_{\rm L}$ is invariant. The phase-frame locking enters through the invariant combination ${\cal V}_\mu$.
With this rotated Nambu spinor relabeled as $\Psi_{\rm L}$, the leading left-handed nodal Lagrangian takes the form
\begin{equation}
  {\cal L}_{\rm node,L}
  ={\cal L}^{(0)}_{\rm node,L}
  -\Psi_{\rm L}^\dagger\left(\tau_3^{\rm N}{\cal V}_0+\ell^i{\cal V}_i\right)\Psi_{\rm L}+\cdots,
  \label{eq:nodal-BdG-texture}
\end{equation}
where ${\cal L}^{(0)}_{\rm node,L}$ is understood with the local-triad substitutions above.
Collecting the different low-energy sectors, the IR theory may be written as
\begin{equation}
  S_{\rm IR} = \int \dd^4x\,\left[{\cal L}_{\rm node,L}+ {\cal L}_{\rm node,R} + {\cal L}_{\rm boson} \right] + S_{\rm WZ} +\cdots.
  \label{eq:EFT}
\end{equation}
The WZ term $S_{\rm WZ}$ reproduces the perturbative mixed anomaly, while the nodal quasiparticles described by ${\cal L}_{{\rm node},\chi}$ carry the global $\mathbb Z_2$ Witten anomalies. 
The topological phase-orientation structure encoded in ${\cal L}_{\rm boson}$ arises from the nonzero angular momentum of the axial Cooper pair and gives rise to the Berry phase and Mermin--Ho response of the orientational NG sector.
Although point-node fermions are subleading in the derivative expansion of a generic spinful-superfluid EFT \cite{Son:2026tcm}, they cannot be discarded in this QCD realization because they carry the Witten anomalies.

\sect{Discussion and outlook}%
In this paper, we studied the flavor-singlet $J^P=1^+$ secondary pairing of the residual blue quarks in the 2SC phase. 
The phenomenological relevance of this phase at neutron-star densities remains uncertain, since the pairing dynamics is not under perturbative control.
The instanton-based NJL model analysis of Ref.~\cite{Alford:2002rz} finds a gap parameter of order $1\,{\rm MeV}$ in this channel at intermediate density. 
Although small compared with the QCD scale, such a gap can still control the low-energy quasiparticle spectrum of the residual blue-quark sector and be relevant to neutron-star phenomenology.
In $\beta$-equilibrated neutral matter, however, the up- and down-quark Fermi-surface mismatch can exceed this scale, disfavoring homogeneous secondary pairing. 
Whether this pairing survives in realistic neutron-star matter remains an important microscopic question.
If realized, the point-node quasiparticles can have direct consequences for the specific heat, transport coefficients, and weak-interaction rates relevant to neutron-star cooling.

The EFT derived here also provides a systematic framework for studying collective and rotational phenomena beyond the quasiparticle sector. In addition to the superfluid phonon with linear dispersion, the axial state supports a type-B orientational NG mode with quadratic dispersion. 
It gives a 3D density of states $\rho(\omega)\propto \omega^{1/2}$, yielding a low-temperature specific heat $C\propto T^{3/2}$, which can parametrically dominate the $T^3$ contributions from the phonon and nodal quasiparticles.  

The phase-orientation coupling may be important for the rotational dynamics of neutron stars. Through the Mermin--Ho relation, smooth textures of the nodal-axis field $\bm\ell$ carry superfluid vorticity, so that rotating matter can in principle support texture-induced vortex configurations in addition to conventional singular vortices. The EFT therefore provides a basis for computing vortex structure, tension, collective excitations, and their coupling to nodal quasiparticles. These ingredients are relevant to mutual friction and the damping of macroscopic stellar oscillations.

\sect{Acknowledgements}%
The author thanks Y.~Fujimoto for useful correspondence. The author is supported by JSPS KAKENHI Grant No.~JP24K00631 and No.~JP26K00698.

\bibliography{2SC}

\section*{End Matter}
\appendix
\section{Dual formulation of the phase-orientation coupling}
\label{app:dual_formulation}

The coupling between the superfluid and orientational sectors admits a dual formulation in terms of a two-form gauge field \cite{Son:2026tcm}. We first recall that the conserved modified-baryon current satisfies $\partial_\mu j_{\widetilde{\rm B}}^\mu=0$. Locally, this conservation law can be solved identically by introducing a two-form gauge field ${\cal B}_{\mu\nu}$, such that $j_{\widetilde{\rm B}}^\mu =\frac12\epsilon^{\mu\nu\lambda\rho} \partial_\nu{\cal B}_{\lambda\rho}$.
This current is invariant under the two-form gauge transformation 
${\cal B}_{\mu\nu} \rightarrow {\cal B}_{\mu\nu}+\partial_\mu\Lambda_\nu-\partial_\nu\Lambda_\mu$.

The orientational field $\bm\ell$ defines the two-form
\begin{equation}
  J_{\mu\nu}[\bm\ell]\coloneqq\epsilon_{abc}\ell^a\partial_\mu\ell^b\partial_\nu\ell^c=\bm\ell\cdot\left(\partial_\mu\bm\ell\times\partial_\nu\bm\ell\right).
  \label{eq:orientation-current}
\end{equation}
Geometrically, $J_{\mu\nu}$ is the pullback of the area form on the order-parameter sphere $S^2$ to spacetime. It is thus closed, $\partial_{[\lambda}J_{\mu\nu]}=0$. Locally, it can be written as the curvature of the frame connection $a_\mu$, $J_{\mu\nu} = \partial_\mu a_\nu -\partial_\nu a_\mu$.

The covariant coupling between the conserved current and the orientational connection,
\begin{equation}
  S_{\rm frame} = s\int \dd^4x\,j_{\widetilde{\rm B}}^\mu a_\mu,
  \label{eq:current-frame}
\end{equation}
can now be rewritten in terms of the dual variable ${\cal B}_{\mu\nu}$. Substituting $j_{\widetilde{\rm B}}^\mu =\frac12
\epsilon^{\mu\nu\lambda\rho} \partial_\nu{\cal B}_{\lambda\rho}$ and integrating by parts gives, up to a boundary term,
\begin{align}
  S_{\rm frame}
  &=\frac{s}{2} \int \dd^4x\,\epsilon^{\mu\nu\lambda\rho}\left(\partial_\nu{\cal B}_{\lambda\rho}\right)a_\mu
  \nonumber\\
  &=\frac{s}{4} \int \dd^4x\,\epsilon^{\mu\nu\lambda\rho}{\cal B}_{\mu\nu}\left(\partial_\lambda a_\rho-\partial_\rho a_\lambda \right)
  \nonumber\\
  &=\frac{s}{4} \int \dd^4x\,\epsilon^{\mu\nu\lambda\rho}{\cal B}_{\mu\nu}J_{\lambda\rho}[\bm\ell].
  \label{eq:dual-topological}
\end{align}
Equation~\eqref{eq:dual-topological} describes a topological interaction of the same form as that in Ref.~\cite{Son:2026tcm}. For the axial 2SC state considered here, the coefficient $s=1/2$ is the angular momentum per unit modified baryon charge. 

\end{document}